\documentclass[aps,prl,twocolumn,superscriptaddress,showpacs,floatfix]{revtex4-1}
\usepackage{amsmath,amssymb,graphicx}
\usepackage{wasysym}
\usepackage[utf8]{inputenc}
\usepackage[T1]{fontenc}
\usepackage{xcolor}

\IfFileExists{newtxtext.sty}
   {\usepackage{newtxtext,newtxmath}}
   {\IfFileExists{stix.sty}
      {\usepackage{stix}}
      {\IfFileExists{mathptmx.sty}
      {\usepackage{mathptmx}}{} } }

\usepackage{textcomp}

\usepackage{bm}

\IfFileExists{siunitx.sty}{\usepackage{booktabs,siunitx}}{}

\pdfoutput=1
\usepackage{color}
\definecolor{LinkColor}{rgb}{0.256,0.439,0.588}
\usepackage{hyperref}

\usepackage{pifont}

\begin{document}

\title{Quantum Spin Liquid with Even Ising Gauge Field Structure on the Kagome Lattice}

\author{Yan-Cheng Wang}
\affiliation{Beijing National Laboratory of Condensed Matter Physics and Institute of Physics, Chinese Academy of Sciences, Beijing 100190, People's Republic of China}
\affiliation{School of Physical Science and Technology, China University of Mining and Technology, Xuzhou Jiangsu, 221116, People's Republic of China}
\author{Xue-Feng Zhang}
\email{Corresponding author: zhangxf@cqu.edu.cn}
\affiliation{Department of Physics, Chongqing University, Chongqing 401331, People's Republic of China}
\affiliation{Max-Planck-Institute for the Physics of Complex Systems, 01187 Dresden, Germany}
\author{Frank Pollmann}
\affiliation{Department of Physics, Technical University of Munich, 85748 Garching, Germany}
\affiliation{Max-Planck-Institute for the Physics of Complex Systems, 01187 Dresden, Germany}
\author{Meng Cheng}
\affiliation{Department of Physics, Yale University, New Haven, Connecticut 06520-8120, USA}
\author{Zi Yang Meng}
\email{Corresponding author: zymeng@iphy.ac.cn}
\affiliation{Beijing National Laboratory of Condensed Matter Physics and Institute of Physics, Chinese Academy of Sciences, Beijing 100190, People's Republic of China}
\affiliation{CAS Center of Excellence in Topological Quantum Computation and School of Physical Sciences, University of Chinese Academy of Sciences, Beijing 100190, People's Republic of China}

\begin{abstract}
	Employing large-scale quantum Monte Carlo simulations, we study the extended $XXZ$ model on the kagome lattice. A $\mathbb Z_2$ quantum spin liquid phase with effective even Ising gauge field structure emerges from the delicate balance among three symmetry-breaking phases including stripe solid, staggered solid and ferromagnet. This $\mathbb{Z}_2$ spin liquid is stabilized by an extended interaction related to the Rokhsar-Kivelson potential in the quantum dimer model limit. The phase transitions from the staggered solid to a spin liquid or ferromagnet are found to be first order and so is the transition between the stripe solid and ferromagnet. However, the transition between a spin liquid and ferromagnet is found to be continuous and belongs to the 3D $XY^*$ universality class associated with the condensation of spinons. The transition between a spin liquid and stripe solid appears to be continuous and associated with the condensation of visons.
\end{abstract}

\date{\today}

\maketitle

\paragraph*{Introduction.} Quantum spin liquids (QSLs)~\cite{BalentsSLReview,SLreview,LeonReview17} are representatives of topologically ordered states of matter~\cite{Wen2016}, characterized by long-range many-body entanglement and fractionalized excitations. In the zoo of QSLs, the $\mathbb Z_2$ spin liquid, whose elementary excitations are coupled to the emergent Ising gauge field~\cite{WenZ2SL1991,Wen1991a}, can be realized in an extended $XXZ$ spin model on a kagome lattice, i.e., the Balents-Fisher-Girvin (BFG) model~\cite{BFG2002,Isakov2006,Isakov2011,Isakov2012,YCWang2017}, which has been extensively investigated as one of the few models of frustrated magnets that can be simulated with unbiased quantum Monte Carlo (QMC) methods.
There are two promising QSL materials with kagome lattice geometry -- ZnCu$_{3}$(OH)$_{6}$Cl$_2$ (Herbertsmithite)~\cite{Fu2015,Han2016} and Cu$_{3}$Zn(OH)$_{6}$FBr (Zn-doped Barlowite)~\cite{FengZL17,WenXG17,WeiYuan2017,Feng1712,Pasco2018}. In both cases, several experiments are pointing towards gapped QSL ground states with possibly $\mathbb Z_2$ topological order~\cite{FengZL17,ZaletelPRL2015}, especially the latter, in which a gapped spinon continuum has been consistently revealed from both nuclear magnetic resonance~\cite{FengZL17} and inelastic neutron scattering experiments~\cite{WeiYuan2017}, and phase transitions from a magnetically ordered phase to the QSL are realized by tuning the chemical substitution~\cite{Feng1712,Pasco2018}. Therefore, controlled theoretical investigations that could shine light on the transition from $\mathbb Z_2$ topological order to a symmetry breaking phase would be very useful to further guide the experiment developments.

\begin{figure}[htp!]
	\centering
	\includegraphics[width=\columnwidth]{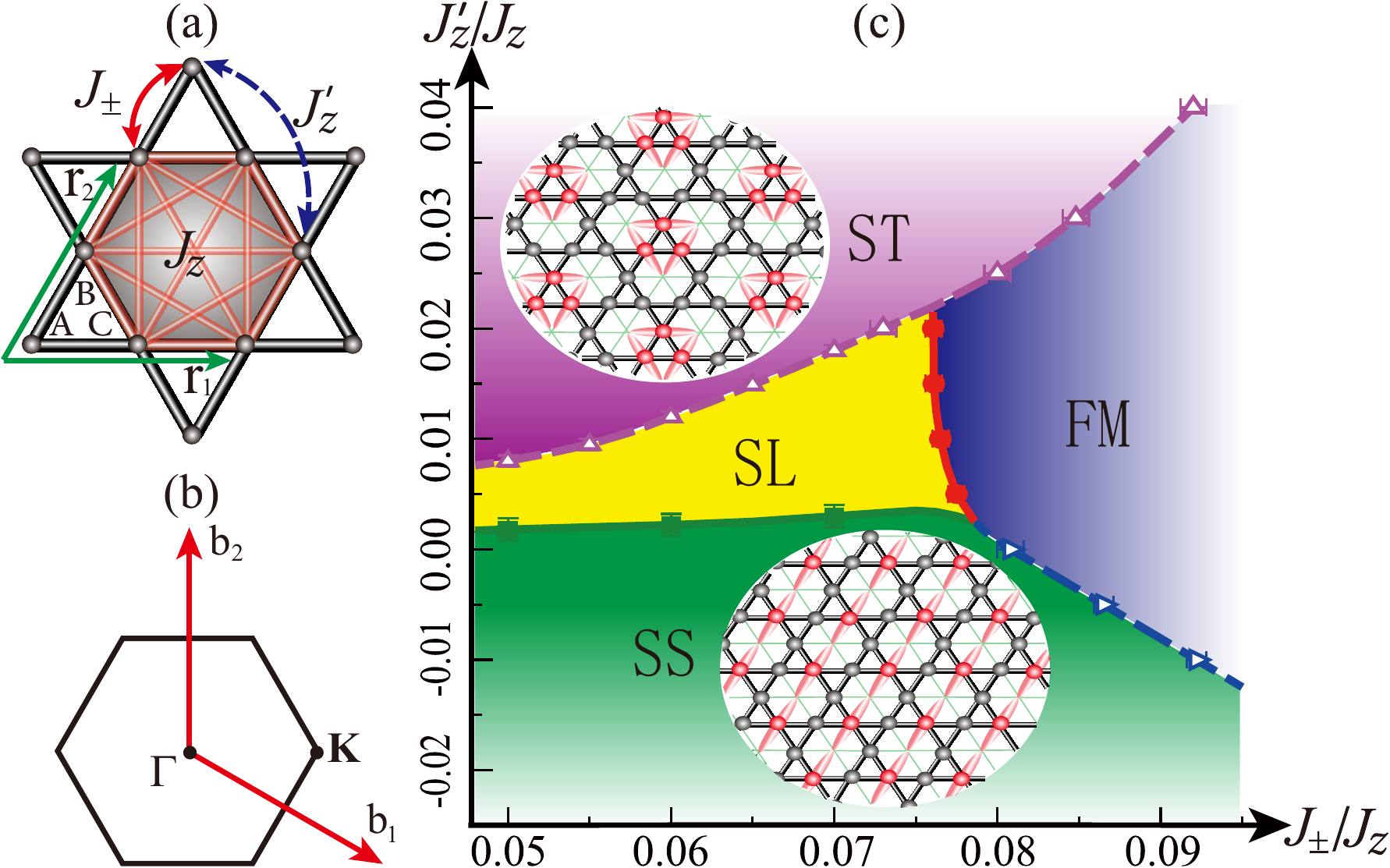}
	\caption{(a) The kagome lattice with lattice vectors $\mathbf{r}_1$ and $\mathbf{r}_2$, sublattices $A$, $B$ and $C$, and all the interactions $J_{\pm}$, $J_z$, and $J'_z$ of the Hamiltonian in Eq.~(\ref{eq:hamiltonian}) are depicted. (b) Brillouin zone of the kagome lattice, with the reciprocal space vectors $\mathbf{b}_1$ and $\mathbf{b}_2$, the high-symmetry points $\Gamma=(0,0)$, and $K=(4\pi/3,0)$.  (c) $J_{\pm}/J_z$-$J'_z/J_z$ phase diagram of the Hamiltonian in Eq.~(\ref{eq:hamiltonian}) with magnetization $m^{z}=1/6$ (or $1/3$ filling factor in boson language). The yellow, green, violet, and blue shaded areas are $\mathbb Z_2$ spin liquid (SL), stripe solid (SS), staggered solid (ST), and ferromagnetic (FM) phases, respectively. The arrangements of the spin configuration in the SS and ST phases are schematically shown in the insets, with the red (gray)  ball denoting spin up (down). The red dimers stand for the effective dimer covering in the SS and ST phases. The phase transition SL-FM is continuous, SL-SS is seemingly continuous, and SL-ST, ST-FM, and SS-FM are first order.}
	\label{fig:LattPHD}
\end{figure}

Theoretically, the ground state phase diagram of a spin-1/2 model is constrained by the celebrated Lieb-Schultz-Mattis-Oshikawa-Hastings (LSMOH) theorem, which asserts that, for systems with an odd number of spin 1/2 per unit cell (or fractional filling of bosons per unit cell), any trivial gapped ground state is forbidden~\cite{LSM, OshikawaLSM, HastingsLSM}. As a result, the phase diagram should contain only symmetry-breaking phases  or spin liquids. This is indeed the case for the BFG model at a zero external field, whose phase diagram consists of a ferromagnet and a gapped $\mathbb{Z}_2$ spin liquid~\cite{YCWang2017}. A further refinement of the LSMOH theorem~\cite{ChengLSM, ZaletelPRL2015} implies that the $\mathbb Z_2$ spin liquid, which can be viewed as an Ising gauge theory, must have an effective odd gauge structure. Intuitively, the ground state contains one Ising gauge charge per unit cell due to Gauss's law. This is a manifestation of the nontrivial fractionalization of lattice symmetries in this phase~\cite{Essin2013,Barkeshli2014,Tarantino2016, Chen2014}, that the anyon excitations have fractional lattice momentum. This fact has important implications for lattice symmetry-breaking phases proximate to the spin liquid.

%As for the BFG model, at zero external field (or half-filling in the hard-core boson language), the ground state phase diagram has been mapped out via QMC~\cite{Isakov2006,Isakov2011,Isakov2012,YCWang2017}. According to the celebrated Lieb-Schultz-Mattis-Oshikawa-Hastings (LSMOH) theorem, for systems with an odd number of spin-1/2 per unit cell (or fractional filling of bosons per unit cell), any trivial gapped ground state is forbidden~\cite{LSM, OshikawaLSM, HastingsLSM}, so the phase diagram contains only symmetry-breaking phases (i.e. ferromagnet or superfluid in the boson language) and a gapped $\mathbb Z_2$ spin liquid~\cite{YCWang2017}. A further refinement of the LSMOH theorem~\cite{ChengLSM, ZaletelPRL2015} implies that the $\mathbb Z_2$ spin liquid must have an effective odd Ising gauge structure, a manifestation of the non-trivial fractionalization of lattice symmetries in this phase~\cite{Essin2013,Barkeshli2014,Tarantino2016, Chen2014}.
%In the strong coupling limit, the $\mathbb Z_2$ spin liquid phase can be understood explicitly from a mapping to quantum dimer model on triangular lattice (spin up mapped to dimers sitting on corresponding bonds). Because an odd number of dimers originates from each hexagon of the kagome lattice, an odd Ising gauge field structure~\cite{Wegner1971,Fradkin1978,Kogut1979,Moessner2001} is realized.

On the other hand, if we turn to one-third filling (integer bosons per unit cell) in the BFG model, which is outside the realm of the LSMOH theorem, a $\mathbb Z_2$ spin liquid ground state with an even Ising gauge field structure may exist (a featureless Mott insulator is also possible), and, if so, its emergent anyon excitations will host different fractional quantum numbers and fractionalization patterns~\cite{ChengLSM,GYSun2018}. In addition, the absence of the LSMOH constraint implies that more symmetry-breaking phases may compete with the potential $\mathbb Z_2$ spin liquid. We thus expect to see a richer phase diagram, with possibly more phase transitions of exotic type driven by condensation of fractionalized anyonic excitations.
%Although the existence of such a $\mathbb{Z}_2$ QSL in the BFG model at one third filling remains controversial~\cite{Roychowdhury2015,Plat2015}, the overall the situation is actually more similar to the experimental discoveries of the Cu$_{4-x}$Zn$_{x}$(OH)$_{6}$FBr (Zn-doped Barlowite) familiy, whose magnetic structure could host several different symmetry-breaking patterns~\cite{Feng1712,Pasco2018,Ranjith2018}. Therefore, it is timely both from theoretical and experimental perspectives that, in this work, we provide that a generalized BFG model that can stabilize this new type of $\mathbb{Z}_2$ QSL and its phase transition to different symmetry-breaking phases.

\paragraph*{Model and Method.} In this Letter, we study a system hosting a $\mathbb Z_2$ spin liquid with an even Insing gauge field structure and solve it with large scale QMC simulations. The Hamiltonian of the extended $XXZ$ model on the kagome lattice is given by
\begin{equation}
	\begin{split}
		H =&-J_{\pm}\sum_{\langle i,j \rangle} (S_i^{+}S_j^{-}+\text{H.c.}) + \frac{J_z}{2}\sum_{\hexagon}\Big(\sum_{i\in\hexagon} S_i^z\Big)^{2}\\
 &+J'_z\sum_{\langle i,j \rangle^{'}} S_i^{z}S_j^{z}-h\sum_{i}S_i^{z},
	\end{split}
\label{eq:hamiltonian}
\end{equation}
where the physical meaning of each term is illustrated in Fig.~\ref{fig:LattPHD} (a). The original BFG model consists of the nearest-neighbor spin flip $J_{\pm}>0$ term and $J_z>0$ plaquette interaction terms for each hexagon which induces local degeneracy~\cite{BFG2002}. The newly added $J'_z$ term is a  next-nearest-neighbor interaction that frustrates the ordering of spins on the same sublattice. The Zeeman field $h$ here is used to tune the magnetization. Throughout the Letter, we set $J_z=1$ as the energy unit.

\begin{figure}[htp!]
	\centering
	\includegraphics[width=\columnwidth]{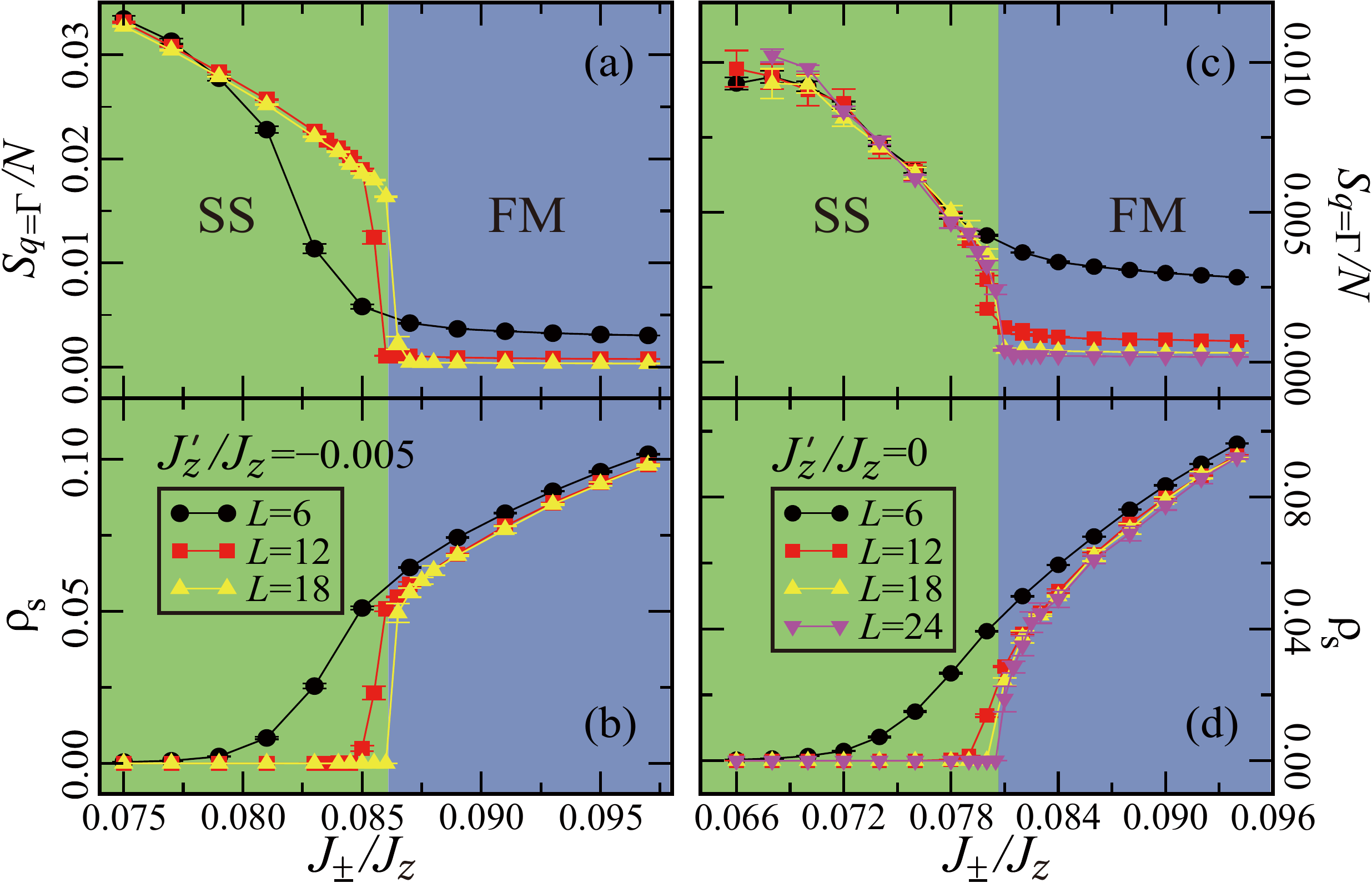}
	\caption{The structure factor $S_{\mathbf{q}=\Gamma}/N$ and spin stiffness $\rho_s$ of the system  as a function of   $J_{\pm}/J_z$ at $J'_{z}/J_{z}=-0.005$ [$S_{\mathbf{q}=\Gamma}/N$ (a) and $\rho_s$ (b)] and  $J'_{z}/J_z=0 $ [$S_{\mathbf{q}=\Gamma}/N$ (c) and $\rho_s$ (d)]. The inverse temperature is set to $\beta J_{\pm}=2L$, and the initial spin configuration is set to be  inside the SS phase.}
	\label{fig:SSFM}
\end{figure}

\begin{figure}[htp!]
	\centering
	\includegraphics[width=\columnwidth]{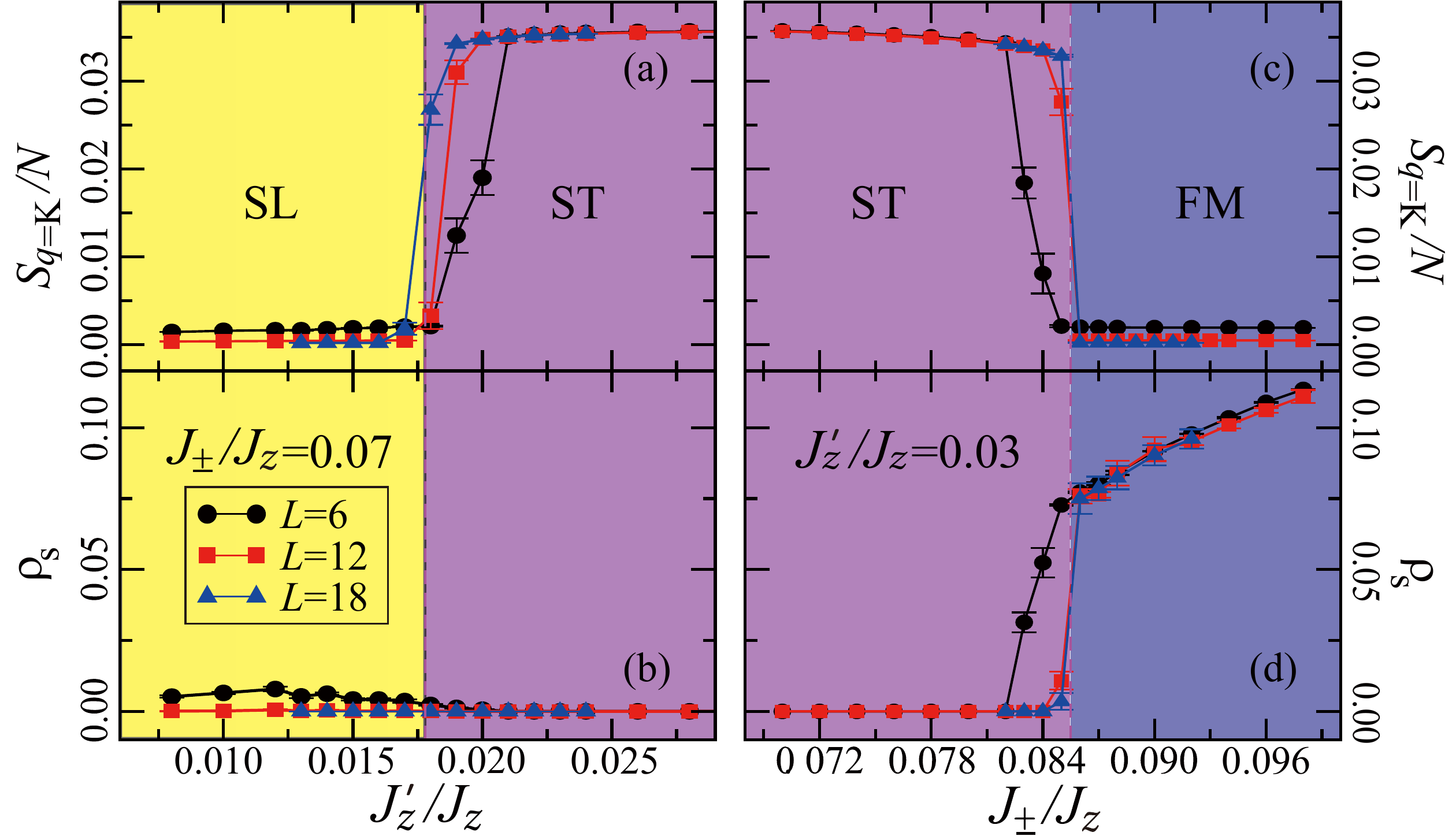}
	\caption{The structure factor $S_{\mathbf{q}=K}/N$ and spin stiffness $\rho_s$ of the system  as a function of  $J'_{z}/J_z$ at $J_{\pm}/J_z=0.07$ [$S_{\mathbf{q}=K}/N$ (a) and $\rho_s$ (b)] and as a function of $J_{\pm}/J_z$  at $J'_{z}/J_z=0.03$ [$S_{\mathbf{q}=K}/N$ (c) and $\rho_s$ (d)]. The inverse temperature is set to $\beta J_{\pm}=2L$, and the initial spin configuration is set to be  inside the ST phase.}
	\label{fig:sfsq-ST}
\end{figure}

We note that, as shown in previous work~\cite{YCWang2017}, the $\mathbb Z_2$ spin liquid in the BFG model (i.e., $J_z'=0$) can be stabilized when the magnetization is zero, i.e., $m^{z}=\frac{1}{6}\sum_{i\in\hexagon}S_i^z=0$. As we have mentioned, this filling immediately implies that the $\mathbb Z_2$ spin liquid has an odd Ising gauge structure. In the Ising limit $J_z\gg J_\pm$, where a mapping to the quantum dimer model becomes plausible, $m^{z}=0$ means that three dimers originate from the center of the hexagon. To have an even number of dimers required by the even Ising gauge structure~\cite{Moessner2001}, the net magnetization must be adjusted to 1 on each hexagon, corresponding to $m^{z}=\pm 1/6$. With such a net magnetization, the ground state of the BFG model turns out to be ferromagnetic for large $J_{\pm}/J_z$ but may be a stripe solid (SS) phase~\cite{Plat2015} or spin liquid (SL) phase~\cite{Roychowdhury2015} in strong coupling region $J_{\pm}\ll J_z$. In order to stabilize the SL phase, a diagonal Rokhsar-Kivelson (RK) potential $V_{\text{RK}}$ defined on the corner shared triangles is introduced, and the critical point between SL and the accompanying staggered solid (ST) phase is exactly the RK point $V_{\text{RK}}=4J_{\pm}^2/J_z$. As mentioned in Ref.~\cite{Plat2015}, the effective RK interaction can be inserted as the $J'_z$ term in the original BFG model. As shown with QMC simulations below, it plays a key role in stabilizing the $\mathbb Z_2$ spin liquid with an even Ising gauge field structure in our model.

To reveal the ground state phase diagram of Eq.~(\ref{eq:hamiltonian}), we employ large-scale stochastic series expansion QMC simulations, with a plaquette update~\cite{Syljuaasen2002,YCWang2017} and generalized balance condition~\cite{SuwaPRL2010,YCWang2017}. Since the model is highly anisotropic and frustrated, i.e., $J_{\pm} \ll J_z$ and $J'_{z}$, to avoid the sampling problem of many local minima, we perform QMC updates with a five-spin plaquette update (ten legs in a vortex)~\cite{Melko2007}, instead of the conventional two-spin bond update. Moreover, to reduce the rejection rate of the proposed spin configuration, we make use of an algorithm that satisfies the balance condition without imposing detail balance between the Monte Carlo configurations~\cite{SuwaPRL2010}. Details of our numerical method are presented in the Supplemental Material~\cite{suppl}.

\paragraph*{Results.}
Applying large-scale QMC simulations in canonical ensembles with exact $m^{z}=1/6$, we obtain the phase diagram shown in Fig.~\ref{fig:LattPHD} (c).
Four different phases are distinguished by two physical observables: (i) the spin stiffness (superfluid density in hard-core boson language) $\rho_s=(W_{\mathbf{r}_1}^2+W_{\mathbf{r}_2}^2)/(4\beta J_{\pm})$ through winding number fluctuations $W_{\mathbf{r}_{1,2}}^2$~\cite{Pollock1987}, where $\mathbf{r}_{1,2}$ are the two lattice directions, as shown in Fig.~\ref{fig:LattPHD} (a), and (ii) the sublattice magnetic structure factor $S_\mathbf{q}=\frac{1}{N}\sum_{\{i,j\}}e^{i{\bf r}_{ij}\cdot{\bf q}}(\langle S_i^z S_j^z \rangle-\langle S_i^z \rangle \langle S_j^z \rangle)$ where $\{i,j\}$ represents $i$th and $j$th sites are in the same sublattice, and $N=3L^2$ is the volume of the system. For weak coupling $J_{\pm}\gg J_z$, the spin stiffness converges to a finite value, and the ferromagnetic (FM) order is formed.
At small $J_{\pm}/J_z$, the frustrations induced via $J'_z$ and $J_z$ manifest themselves in an intriguing way, such that three different phases emerge. For  large antiferromagnetic $J'_{z}\gg J_z$, $\rho_s$ vanishes, but the magnetic structure factor $S_{\mathbf{q}}/N$ has a peak at the wave vector $\mathbf{q}=K$ so that the system forms an ST phase, with the bosons (or $S^z_i$) arranging themselves according to a $\sqrt{3}\times \sqrt{3}$ unit cell, as shown in the upper inset in Fig.~\ref{fig:LattPHD} (c). In comparison, for large ferromagnetic interaction $-J'_{z}\gg J_z$, the system is still incompressible with zero $\rho_s$, but $S_{\mathbf{q}}/N$ peaks at the wave vector $\mathbf{q}=\Gamma$, which implies that the system forms another kind of crystalline order [see the lower inset of Fig.~\ref{fig:LattPHD} (c)].
We identify this phase as the SS phase, because the magnetization pattern only breaks the rotation symmetry of the kagome lattice, and it is therefore also a nematic phase~\cite{Roychowdhury2015}. Between these two solid phases, we find a phase without any obvious symmetry breaking, and, as will be explained below, numerical data suggest the existence of fractionalized excitations in this symmetric phase and their condensation transitions into other symmetry-breaking phases. These results, combined with the quantum dimer model limit of our model~\cite{Roychowdhury2015,Plat2015}, suggest that this phase is a $\mathbb Z_2$ SL phase with an even Ising gauge structure.

According to the Ginzburg-Landau theory, the phase transition between SS to FM should be first order, because they break different symmetries; otherwise, an exotic scenario such as a deconfined quantum critical point~\cite{Senthil2004,Qin201705,XFZhang2017,NSMa2018} must be required. In Figs.~\ref{fig:SSFM}(a) and \ref{fig:SSFM}(b), finite structure factor $S_{\mathbf{q}=\Gamma}/N$ and spin stiffness $\rho_{s}$ clearly demarcate the region of SS and FM at large negative $J'_{z}/J_z=-0.005$, respectively. A sharp discontinuity indicates a first-order transition. However, the results at $J'_{z}/J_z=0$ are more subtle, as shown in Figs.~\ref{fig:SSFM}(c) and \ref{fig:SSFM}(d). At first glance, both $S_{\mathbf{q}=\Gamma}/N$ and $\rho_s$ change continuously. But when the system size increases, the jumps in $S_{\mathbf{q}=\Gamma}/N$ and $\rho_s$ become more visible. It hints at a weakly first-order transition and may result from the energy gap shrinking when approaching the tricritical point among SL, SS and FM phases. Furthermore, a smaller finite-size effect is observed in the SS phase, which suggests that the SS phase is more stable in the original BFG model, compared with the SL phase.

\begin{figure}[tp!]
	\centering
	\includegraphics[width=\columnwidth]{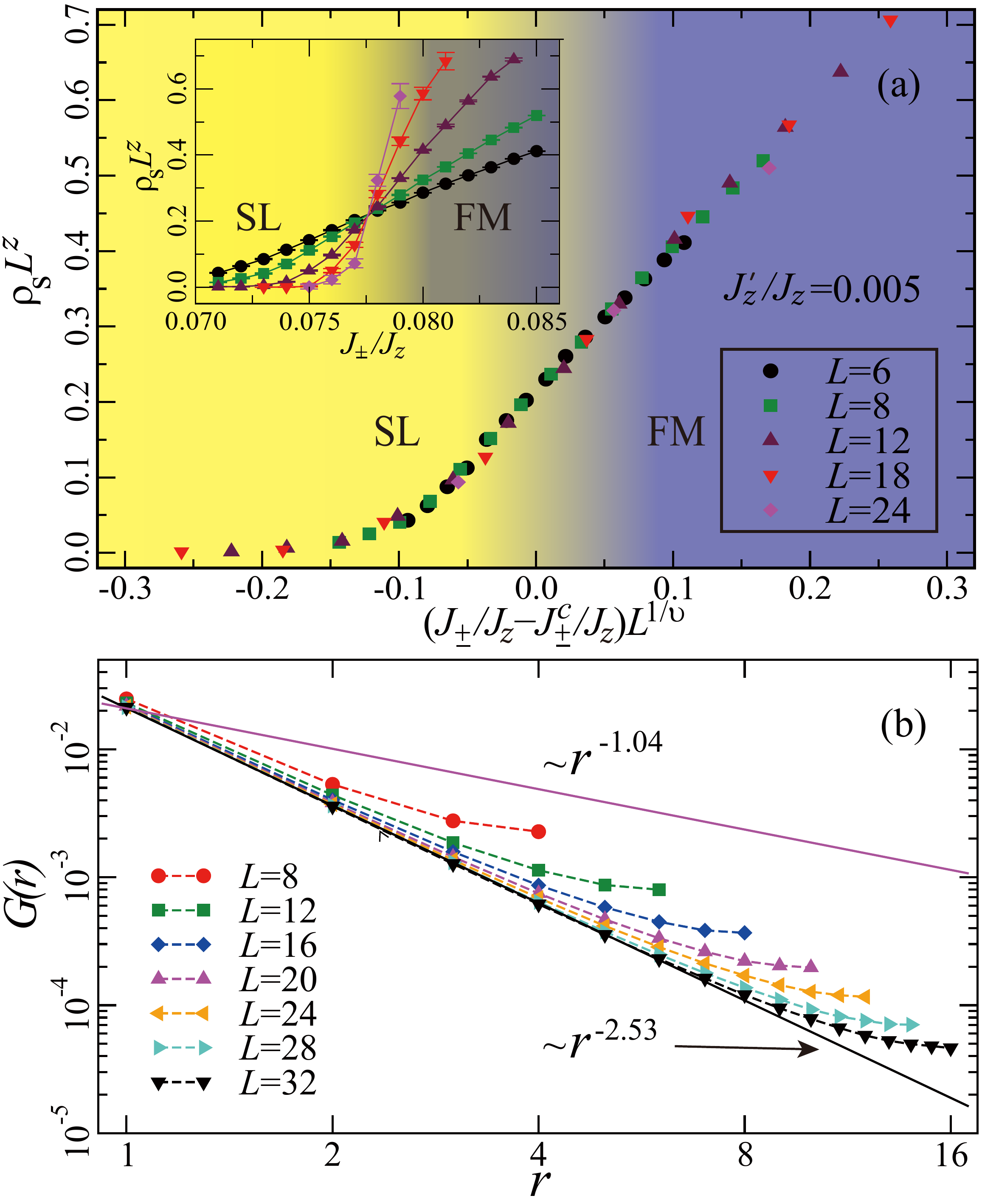}
	\caption{Data collapse of the spin stiffness $\rho_s L^z$ as a function of   $(J_{\pm}/J_z-J^{c}_{\pm}/J_z)L^{1/\nu}$ at (a) $J'_{z}/J_{z}=0.005$ with 3D $XY$ exponents. The inverse temperature is set to $\beta J_{\pm}=2L$. The critical point is determined as $J_{\pm}^c/J_{z}=0.0775$; see the inset. (b) Equal-time spin-spin correlation function $G(r)=\langle S_0^+S_{r}^-\rangle$  as a function of the distance $r$  at critical point $J_{\pm}^{c}/J_z=0.0775$ for $J'_{z}/J_z=0.005$. The log-log plot gives rise to $G(r)\sim r^{-(1+\eta)}=r^{-2.53(4)}$, and large anomalous dimension $\eta=1.53(4)$ is consistent with the 3D $XY^{*}$ transition; the $r^{-1.04}$ line stands for the conventional 3D $XY$ behavior with $\eta=0.04$.}
	\label{fig:rhos_gr}
\end{figure}

Similarly, the phase transitions from ST to FM is first order. As shown in the right part in Fig.~\ref{fig:sfsq-ST}, different from SS, the magnetic structure factor $S_{\mathbf{q}}/N$ has a finite value at $\mathbf{q}=K$ in the ST phase. With fixed $J'_z/J_z=0.03$ and increasing $J_{\pm}/J_z$ up to around $J_{\pm}/J_z \sim 0.085$, $S_{\mathbf{q}=K}/N$ drops to zero and $\rho_s$ jumps to a finite value. Then, the system enters the FM phase through a first-order phase transition.

Surrounded by three symmetry-breaking phases, the SL is located in the middle of the phase diagram. In Fig.~\ref{fig:sfsq-ST}(a) and \ref{fig:sfsq-ST}(b), both order parameters are zero, which means the SL phase does not break related symmetries. Drawing lessons from the quantum dimer model on the triangular lattice~\cite{MoessnerFFIM2001}, the phase transition between SL and ST is expected to be first order around the RK point $J'_z=4J^2_{\pm}/J_z$, where SL and ST are degenerate. From the numerical result shown in Fig.~\ref{fig:sfsq-ST}(a) and \ref{fig:sfsq-ST}(b) with fixing $J_{\pm}/J_z=0.07$, we can find a clear first-order  transition at $J'_z/J_z\sim 0.0175$ matching well with the analytic result from the quantum  dimer model. To further make sure there is no long-range order developed in the SL phase, we plot the S$_{\mathbf{q}}$ for the entire Brillouin zone in Fig.~\ref{fig:Sq_SL} (see Supplemental Material) with $J_{\pm}/J_z=0.07$ and $J'_z/J_z=0.01$ and no obvious Bragg peak is observed.

The intrinsic characteristics of SL are fractionalized topological excitations. In a $\mathbb Z_2$ SL, spinons and visons are deconfined. When quantum fluctuations are enhanced, spinons can condense to form a FM phase, but the universality class of transition is usually special~\cite{Isakov2012}. In order to check it, we implement finite-size scaling on the spin stiffness $\rho_s$ at $J'_z/J_z=0.005$. Seen from the inset in Fig.~\ref{fig:rhos_gr} (a), $\rho_s$ is a continuous function across the SL to FM transition for different sizes. Multiplying $L^z$ by the conjectured dynamical exponent $z=1$, we find that all the curves cross at the critical point $J^{c}_{\pm}/J_z=0.0775$. Then, Fig.~\ref{fig:rhos_gr} (a) presents the data collapse of the stiffness $\rho_s L^{z}$ vs $(J_{\pm}/J_z-J^{c}_{\pm}/J_z)L^{1/\nu}$, where $\nu$ is the correlation length exponent of the 3D $XY$ universality class~\cite{Hasenbusch1999,Meng2008}. Apparently, with $z=1$ and $\nu=0.672$, the spin stiffness is well collapsed. Hence, it suggests that the SL to FM transition is a genuine continuous phase transition of the 3D $XY$ universality.

\begin{figure}[tp!]
\centering
\includegraphics[width=\columnwidth]{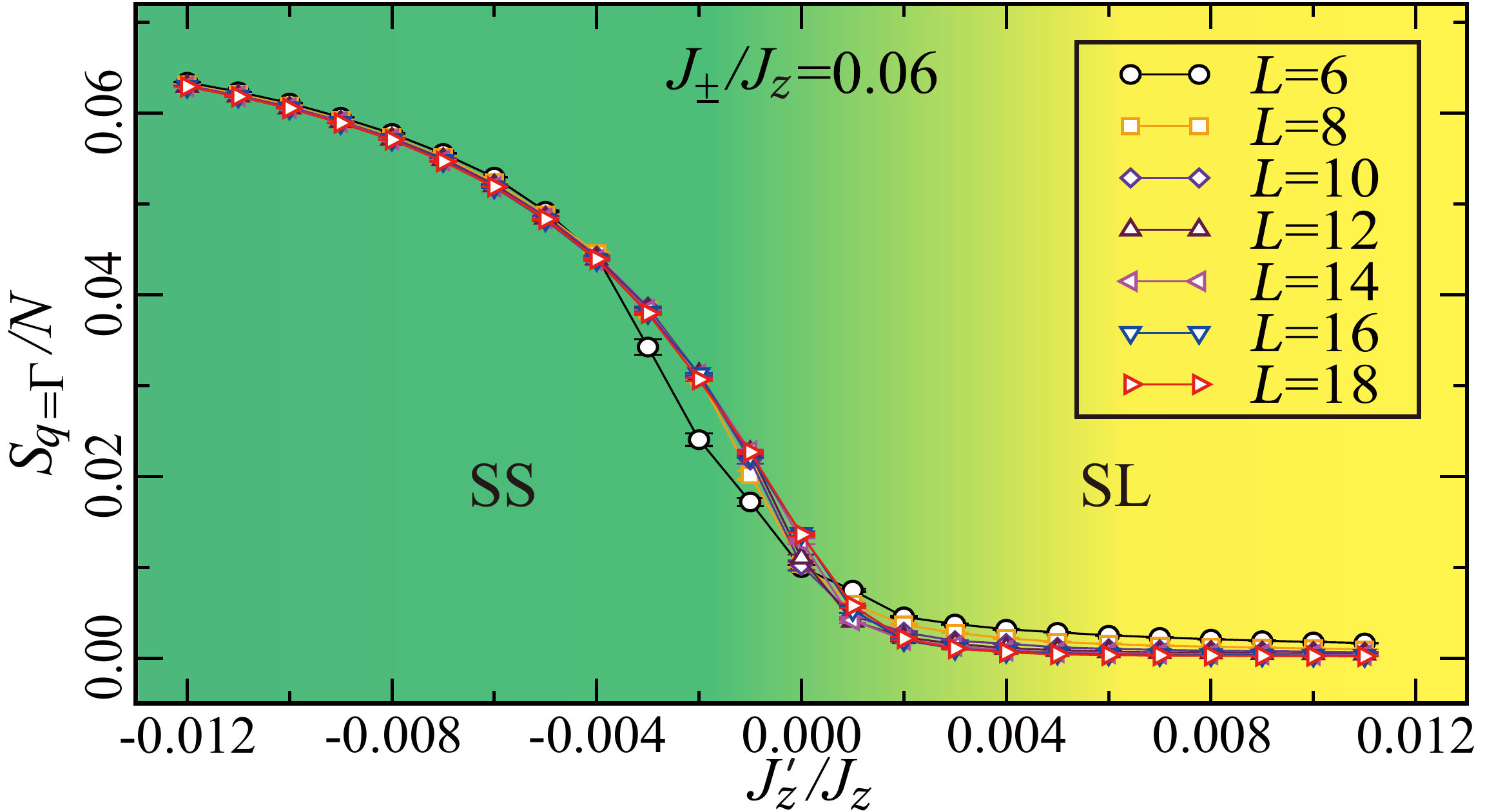}
\caption{The magnetic structure factor $S_{\mathbf{q}=\Gamma}/N$ of the system as a function of $J'_z$ at $J_{\pm}/J_z=0.06$. The inverse temperature is set to $\beta J_{\pm}=2L$, and the initial spin configuration is set to be inside the SS phase.}
\label{fig:sf}
\end{figure}

On the other hand, if the symmetric phase observed at small $J_{\pm}/J_z$ is indeed a $\mathbb Z_2$ SL, the transition from SL to FM should actually belong to the $XY^{*}$ class: Namely, due to the existence of spinon excitations near the ground state, the physical order parameter field $S^+$ fractionalizes into two spinons, which results in a large anomalous dimension $\eta$ at the critical point. This is indeed observed for the transition from the $\mathbb Z_2$ SL to the FM phase at half filling with $\eta=1.5$~\cite{Isakov2012,Kamiya2015}. To see whether such a scenario is realized in our model, we consider the equal-time spin-spin correlation function $G(\mathbf{r})=\langle S^{+}_{0}S^{-}_{\mathbf{r}} \rangle$. At the critical point, $G(\mathbf{r})$ should decay as a power law $G(\mathbf{r})\sim 1/|\mathbf{r}|^{1+\eta}$. In order to minimize the finite-size effect, we plot the $G(\mathbf{r})$ for several different system sizes $L=8, 12, 16, 20, 24, 28$, and $32$ and look for the converged power $1+\eta$ of the real space decay. Figure~\ref{fig:rhos_gr} (b) shows the results at the critical point in a log-log plot, and the converged exponent is found to be $\eta=1.53(4)$ for $J'_z/J_z=0.005$. To distinguish this transition from the conventional 3D XY universality, the $\eta=0.04$ line of the 3D $XY$ behavior is also shown in Fig.~\ref{fig:rhos_gr} (b). These numerical observations confirm the scenario that the transition between SL and FM, as shown in Fig.~\ref{fig:LattPHD} (c), is a 3D $XY^{*}$ transition with a large anomalous scaling dimension, identical with that in the $m^z=0$ ($\mathbb Z_2$ SL with an odd Ising gauge structure) case~\cite{Isakov2006,Isakov2011,Isakov2012,Kamiya2015}; thus, the existence of fractionalized spinon excitations in the SL phase is revealed.

Finally, we consider the transition from SS to SL. Our data of the magnetic structure factor as a function of $J_z'/J_z$ are shown in Fig.~\ref{fig:sf}. Up to $L=18$, $S_{\mathbf{q}=\Gamma}/N$ seems to converge to a continuous curve, i.e., smoothly vanishing as $J_z'/J_z$ goes from negative to positive.  According to Ref.~\cite{Roychowdhury2015}, a continuous transition  from SL to SS with emergent O(3) symmetry is possible due to the even Ising gauge structure in the SL phase, driven by the condensation of visons. However, since we have to simulate the model at very small values of $J_{\pm}/J_z$, the Monte Carlo dynamics becomes extremely slow even with the advanced update scheme exploited in this work; hence, the numerical results obtained are not sufficient to discern the true nature of this transition (we actually performed the data collapse upon the data in Fig.~\ref{fig:sf} but could not obtain meaningful exponents). We leave this task to future work with even more powerful simulation techniques.

\paragraph*{Discussion.} We investigated the ground state phase diagram of an extended BFG model with large-scale QMC simulations, in which ferromagnet, nematic stripe solid, and staggered solid phases and, most importantly, a $\mathbb Z_2$ spin liquid with an even Ising gauge structure are discovered. The phase transitions of SS-FM, ST-FM, and ST-SL are all found to be first order. The phase transitions of SL-FM and SL-SS appear to be continuous. The phase transition from SL and FM is found to fall in the 3D $XY^{*}$ universality class, signaling the fractionalized spinon excitations in the $\mathbb Z_2$ spin liquid. A continuous transition between SL and SS phases is also consistent with the even Ising gauge structure, where the vison excitations of the $\mathbb Z_2$ spin liquid have no fractionalization of the lattice translation symmetries~\footnote{In fact, one can argue that the background charge should be trivial in this model. The even Ising gauge structure implies that the background charge cannot be a spinon. The continuous transition from the $\mathbb Z_2$ SL to the translation-invariant FM phase implies that the bosonic spinon sees no background flux. These two facts together fix the background charge to be trivial. And such expected behavior of vison condensation is verified in Ref.~\cite{GYSun2018}}, and, thus, can drive a condensation transition to a translation-invariant trivial phase~\cite{GYSun2018}. The $\mathbb Z_2$ spin liquid found here, with its even Ising gauge structure, is an outlier of LSMOH-type theorems, and, hence, more competing phases, exemplified by the SS and ST phases, indeed come into play in the phase diagram. The transitions coming out of the SL phase into SS and FM phases provide two possible routes for the anyon condensation. Our results hence broaden the scope of frustrated spin models in which unexpected topological phases could be present.

\paragraph*{Acknowledgement.} We thank valuable discussions with Sylvain Capponi, Chen Fang and Yang Qi. Y.-C.W. and Z.Y.M. acknowledge funding from the Ministry of Science and Technology of China through National Key Research and Development Program under Grant No. 2016YFA0300502, from the key research program of the Chinese Academy of Sciences under Grant No. XDPB0803, and from the National Science Foundation of China under Grant No. 11421092, No. 11574359, and No. 11674370 as well as the National Thousand-Young Talents Program of China. X.-F.Z. acknowledges fundings from Project No. 2018CDQYWL0047 supported by the Fundamental Research Funds for the Central Universities. F.P. acknowledges the support of the DFG Research Unit FOR 1807 through Grants No. PO 1370/2-1,  SFB 1143, and the Nanosystems Initiative Munich (NIM) by the German Excellence Initiative and the support from European Research Council (ERC) under the European Union’s Horizon 2020 research and innovation program through Grant Agreement No. 771537. M.C. is supported by startup funds from Yale University. We thank the Center for Quantum Simulation Sciences in the Institute of Physics, Chinese Academy of Sciences, the Tianhe-1A platform at the National Supercomputer Center in Tianjin, the John von Neumann Institute for Computing (NIC) on the supercomputer JURECA at J\"{u}lich Supercomputing Centre (JSC), the Allianz f\"{u}r Hochleistungsrechnen Rheinland-Pfalz (AHRP), and the Max-Planck Computing and Data Facility (MPCDF) for their technical support and generous allocation of CPU time.

\bibliography{kagome-bfg_1_3_filling}

\newpage
\begin{center}
\textbf{\large Supplemental Material: Quantum Spin Liquid with Even Ising Gauge Field Structure on Kagome Lattice}
\end{center}
%\author{Yan-Cheng Wang}
%\affiliation{Beijing National Laboratory for Condensed Matter Physics and Institute of Physics, Chinese Academy of Sciences, Beijing 100190, China}
%\affiliation{School of Physical Sciences, University of Chinese Academy of Sciences, Beijing 100190, China}
%\author{Zi Yang Meng}
%\affiliation{Beijing National Laboratory for Condensed Matter Physics and Institute of Physics, Chinese Academy of Sciences, Beijing 100190, China}
%\affiliation{School of Physical Sciences, University of Chinese Academy of Sciences, Beijing 100190, China}
%
\setcounter{equation}{0}
\setcounter{figure}{0}
\setcounter{table}{0}
\setcounter{page}{1}
\makeatletter
\renewcommand{\theequation}{S\arabic{equation}}
\renewcommand{\thefigure}{S\arabic{figure}}
\section{Details in Quantum Monte Carlo Method}
\label{sec:QMC}
The Hamiltonian Eq.~(\ref{eq:hamiltonian}) in the main text is written in a physical transparent form, but for the sake of numerical implementation in the SSE-QMC simulation, it is easier to decompose it into summation of operators defined in one plaquette:
\begin{equation}
H = -\sum_{a}\sum_{b}H_{a,b},
\label{eq:hd}
\end{equation}
where $a=1,2$ stands for the two types of vertices. $a=1$ labels the diagonal part, such as operator $S^z_iS^z_j$,  and $a=2$ labels the off-diagonal part, such as spin exchange term $S^{+}_iS^{-}_j$, and $b$ marks the position of the plaquette. Usually, bond ($4$ legs per vortex in SSE language) is chosen as a plaquette in conventional SSE~\cite{Syljuaasen2002}, however, it is shown~\cite{YCWang2017} that for highly frustrated and anisotropic spin systems (like Eq.~(\ref{eq:hamiltonian})), plaquette update is necessary to ensure efficient and ergodicity updating.

\begin{figure}[htp!]
\centering
\includegraphics[width=0.8\columnwidth]{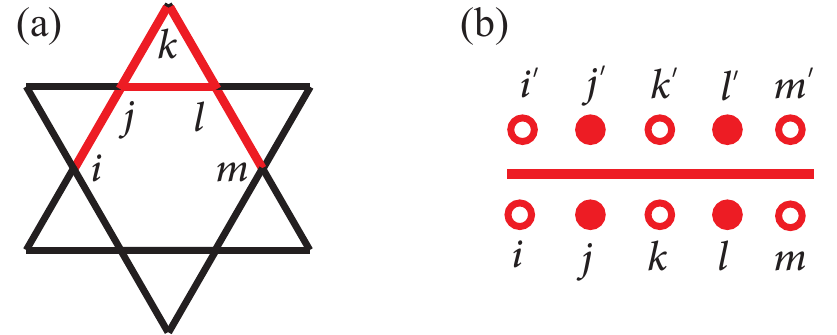}
\caption{(a) A plaquette with $5$ sites $i,j,k,l,m$ chosen for QMC updating. (b) A vortex with $10$ legs in the SSE update.}
\label{fig:vortex}
\end{figure}

Here, as shown in Fig.~\ref{fig:vortex}, we take $5$-site plaquette as a lattice unit during QMC update ($10$ legs in a vortex). Then, the Hamiltonian Eq.~(\ref{eq:hamiltonian}) is decomposed into diagonal operators:
\begin{eqnarray}
H_{1,b} = C & - & \frac{J_z}{z_1} \left(S_i^{z}S_j^{z} +S_j^{z}S_k^{z}+S_k^{z}S_l^{z}+S_j^{z}S_l^{z}+S_l^{z}S_m^{z}\right)\nonumber\\
& - &  \frac{J_z}{z_2}\left(S_i^{z}S_l^{z}+S_j^{z}S_m^{z}\right)-\frac{J_z}{z_2}S_i^{z}S_m^{z}-\frac{J'_z}{z_2}(S_i^{z}S_k^{z}+S_k^{z}S_m^{z})\nonumber\\
& + &  \frac{h}{z_3}\left( S_i^{z}+S_j^{z}+S_k^{z}+S_l^{z}+S_m^{z}\right),
\label{eq:h1b}
\end{eqnarray}
where $z_1=5$, $z_2=2$ and $z_3=10$ are over-counting numbers, and off-diagonal operators:
\begin{equation}
H_{2,b} = \frac{J_{\pm}}{z_1} \left(S_i^{+}S_j^{-} +S_j^{+}S_k^{-}+S_k^{+}S_l^{-}+S_j^{+}S_l^{-}+S_l^{+}S_m^{-} + h.c. \right).
\label{eq:h2b}
\end{equation}
When such large plaquette is selected, we have to face a problem how to get the transfer possibilities between different vortex. Thanks to QMC method without detail balance condition developed by Suwa and Todo~\cite{SuwaPRL2010}, we can easily obtain the solution which strongly reduces the unpreferred bounce update.

\begin{figure}[htp!]
\centering
\includegraphics[width=\columnwidth]{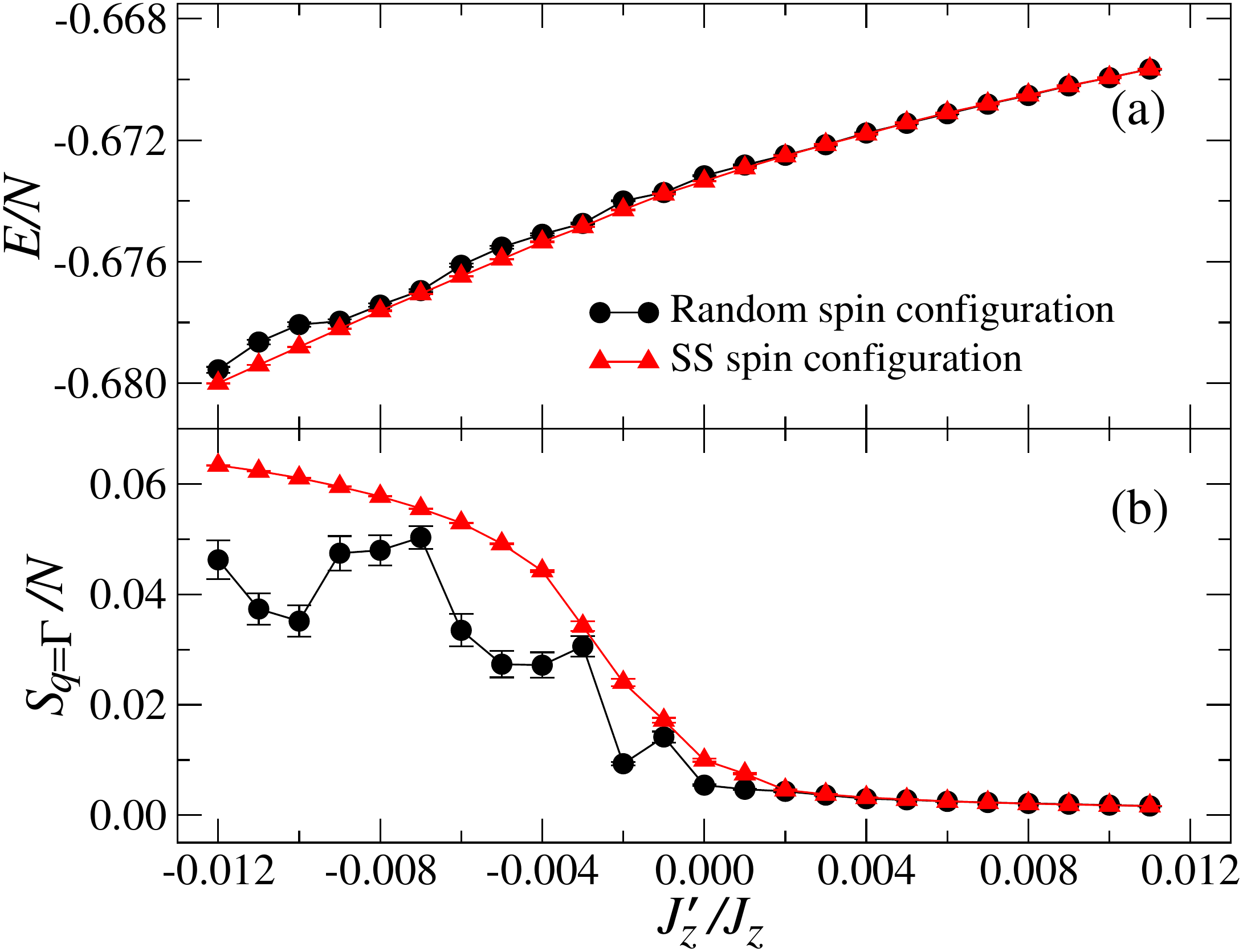}
\caption{ (a) The energy density and (b) the structure factor as a function of $J_z^{'}/J_z$ at $J_{\pm}/J_z=0.06$ for the system size $L=6$ by using two different initial spin configurations: high-T and SS phase. The phase transition is from SS to SL.}
\label{fig:initial_SS}
\end{figure}

\begin{figure}[htp!]
\centering
\includegraphics[width=\columnwidth]{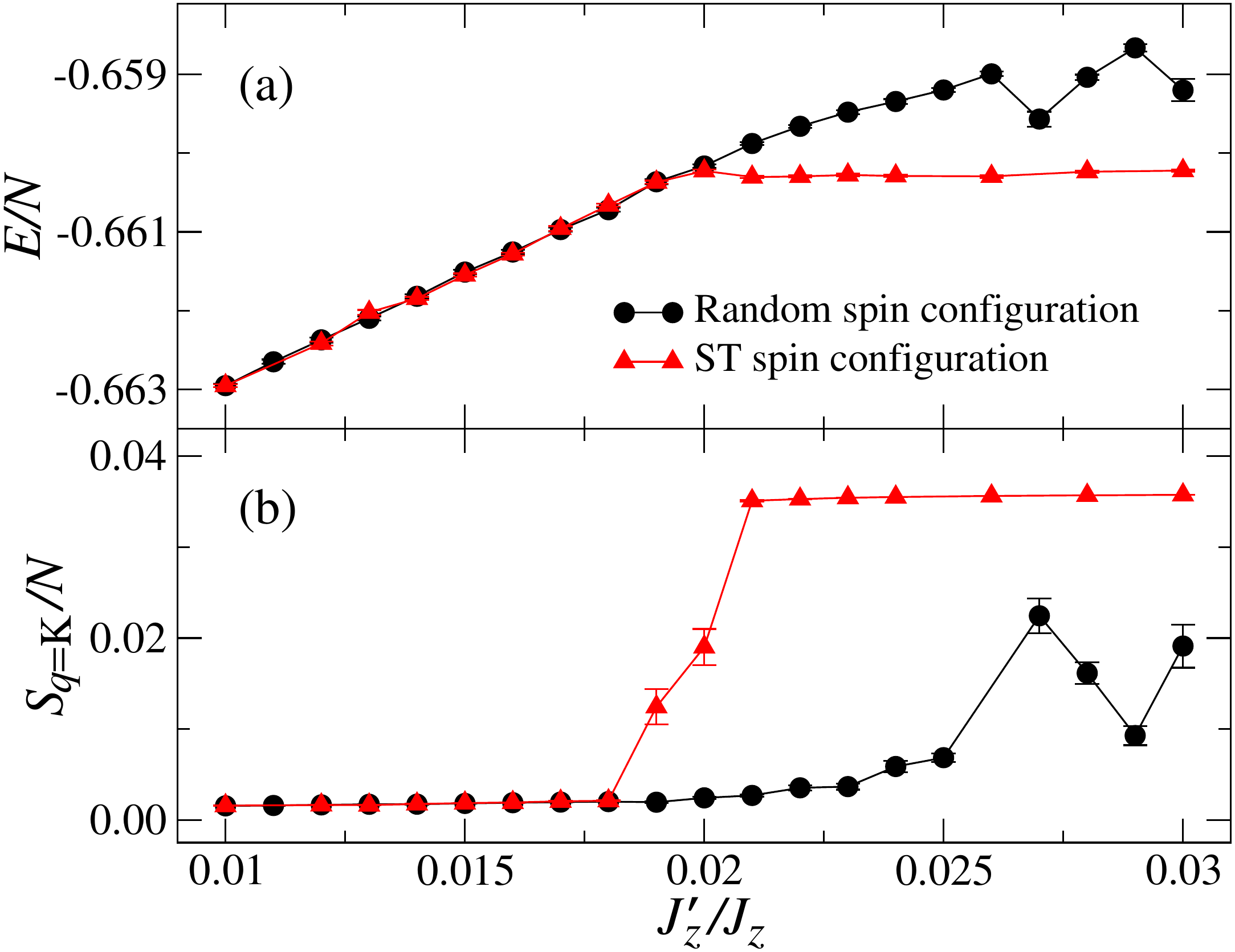}
\caption{ (a) The energy density and (b) the structure factor as a function of $J_z^{'}/J_z$ at $J_{\pm}/J_z=0.07$ for the system size $L=6$ by using two different initial spin configurations: high-T and ST phase. The phase transition is from SL to ST.}
\label{fig:initial_ST}
\end{figure}

Furthermore, to avoid metastable state, we take configuration of SS phase as initial state instead of high temperature (high-T) state (or random spin configuration) when simulating phase transition from SS to FM (or to SL) as shown in Fig.~\ref{fig:SSFM} and Fig.~\ref{fig:sf} in the main text. The comparison of SS phase and high-T initial case is shown in Fig.~\ref{fig:initial_SS}. As we can see, the energy density shown in Fig.~\ref{fig:initial_SS} (a), obtained from SS case is lower than high-T case, especially in the SS phase. Smooth magnetic structure factor with SS initialization in Fig.~\ref{fig:initial_SS} (b) also demonstrates such simulation is more physical and convincing. Similar phenomena can also be found in phase transition between ST and SL reflected in Fig.~\ref{fig:initial_ST}.

\begin{figure}[htp!]
\centering
\includegraphics[width=\columnwidth]{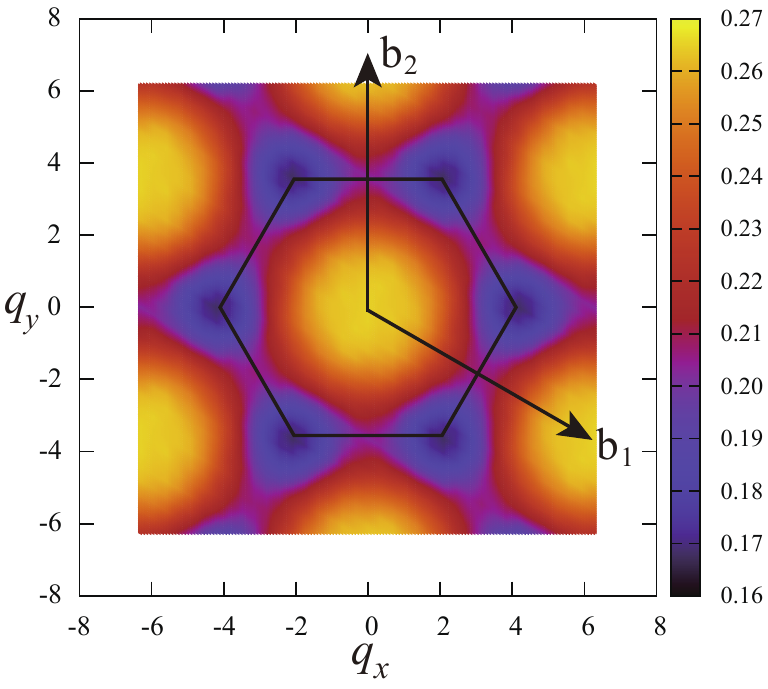}
\caption{Magnetic structure factor $S_\mathbf{q}$ in the entire BZ at $J_{\pm}/J_z=0.07$, $J_z^{'}/J_z=0.01$ for the system size $L=18$ and $\beta J_{\pm}=2L$. No signal of long range order is found.}
\label{fig:Sq_SL}
\end{figure}

To further illustrate that there is no long-range order developed in the spin channels inside the SL phase. We present the spin structure factor in the $S^{z}$ channel,  $S_\mathbf{q}=\frac{1}{N}\sum_{\{i,j\}}e^{i{\bf r}_{ij}\cdot{\bf q}}(\langle S_i^z S_j^z \rangle-\langle S_i^z \rangle \langle S_j^z \rangle)$ where $\{i,j\}$ represents $i$th and $j$th sites are in same sublattice, and $N=3L^2$ is the volume of the system, in the entire BZ at a representive point inside the QLS phase ($J_{\pm}/J_z=0.07$, $J_z^{'}/J_z=0.01$) for the system size $L=18$ and $\beta J_{\pm}=2L$, as shown in Fig.~\ref{fig:Sq_SL}. Since such parameter is inside the QSL phase, there is no sharp peak developed, but only very broad feature close to $\Gamma$.

\end{document}